# Simultaneous TRACERS and THEMIS Observations of Reversed Cusp Ion Dispersions and Dual-Lobe Reconnection

M. Øieroset[1], S. A. Fuselier[2], J. B. Bonnell[1], R. A. Roglans[1], J. S. Halekas[3], R. J. Strangeway[4], T. D. Phan[1], R. G. Gomez[2], S. M. Petrinec[5], K. J. Trattner[6], S. R. Shaver[7], K. A. Goodrich[7], S. A. Henderson[3], S. L. Soni[3], V. Angelopoulos[4], B. L. Burkholder[8], H. Cao[4], L-J. Chen[9], H. K. Connor[9], D. M. Miles[3], A. Moore[3], J. Ng[10], Y. Shen[11]

[1]Space Sciences Laboratory, University of California Berkeley, Berkeley, CA, USA

[2]Southwest Research Institute, San Antonio, TX, USA

[3]Department of Physics and Astronomy, University of Iowa, Iowa City, IA, USA

[4]Department of Earth, Planetary, and Space Sciences, University of California Los Angeles, Los Angeles, CA, USA

[5]Lockheed Martin Advanced Technology Center, Palo Alto, CA, USA.

[6]Laboratory for Atmospheric and Space Physics, University of Colorado, Boulder, CO, USA

[7]Department of Physics and Astronomy, West Virginia University, Morgantown, WV, USA

[8]Goddard Planetary Heliophysics Institute, University of Maryland Baltimore County, Baltimore, MD, USA

[9]Heliophysics Science Division, NASA Goddard Space Flight Center, Greenbelt, MD, USA

[10]Department of Astronomy, University of Maryland, College Park, MD, USA

[11]Department of Physics, University of Alberta, Edmonton, AB, Canada

Corresponding author: Marit Øieroset, email: oieroset@berkeley.edu

Short title: TRACERS and THEMIS lobe reconnection

**Key points:**

- TRACERS-2 observed reversed cusp ion dispersions and sunward convection during two consecutive northern cusp crossings
- THEMIS-D magnetopause measurements provide evidence for dual-lobe reconnection in both hemispheres.
- Similar cusp signatures occur for both northward and $B_X$-dominated IMF conditions

**Abstract**

We present observations from two consecutive TRACERS-2 orbits through the northern low-altitude cusp. During the first crossing, TRACERS-2 observed reversed cusp ion dispersion and sunward convection, consistent with magnetopause reconnection tailward of the cusp during this northward IMF interval. Simultaneous THEMIS-D observations at the equatorial magnetopause show heated magnetosheath plasma captured on closed field lines, with similar particle spectra as in in the low-altitude cusp, indicating that reconnection indeed occurred tailward of the cusp and in both hemispheres. When TRACERS-2 traversed the northern cusp again, 95 minutes later, the IMF was dominated by a negative $B_X$ component. Despite the different IMF conditions, TRACERS-2 recorded nearly the same cusp signatures as before, i.e., reversed ion dispersion and sunward convection. The observations indicate that tailward-of-cusp reconnection can occur for both northward and $B_X$-dominated IMF and that these distinct IMF geometries can produce remarkably similar plasma and field signatures in the low-altitude cusp.

## Plain Language Summary

Magnetic reconnection is a process that allows energy and particles from the solar wind to enter Earth's magnetic environment at its boundary, the magnetopause. Its effects can be monitored by low-altitude spacecraft flying through the cusps, two funnel-shaped regions magnetically connected to the magnetopause. We present observations from two consecutive passes of the TRACERS-2 spacecraft through the low-altitude northern cusp, supported by measurements from the THEMIS mission at the equatorial magnetopause. During the first pass, the interplanetary magnetic field was directed northward, and the observations indicate reconnection occurring tailward of the high-altitude cusp in both hemispheres, allowing magnetosheath plasma to be captured on closed magnetic field lines. When TRACERS-2 crossed the cusp again about 95 minutes later, the interplanetary magnetic field had changed significantly and was dominated by a large radial component. Despite this difference, the cusp signatures were nearly identical. These results show that different solar wind magnetic field configurations can produce similar plasma signatures in the low-altitude cusp.

## 1. Introduction

Magnetic reconnection between the interplanetary magnetic field (IMF) and the geomagnetic field is important for the transfer of energy, mass, and momentum into the magnetosphere. During southward IMF, reconnection occurs near the equatorial magnetopause, resulting in open magnetic field lines convecting first poleward and then tailward, across the polar cap.

When the IMF is northward, reconnection can occur tailward of the cusp, between the magnetosheath and lobe fields (e.g., Dungey, 1961; Gosling et al., 1991; Crooker, 1992; Fuselier et al., 1995; Bosqued et al., 2001; Trattner et al., 2004). When tailward-of-cusp magnetopause reconnection occurs first in one hemisphere and then in the other, magnetosheath plasma can become captured on newly formed closed magnetospheric field lines, leading to a cold and dense magnetospheric boundary layer at the dayside magnetopause (Song and Russell, 1992; Raeder et al., 1995; Sandholt et al., 1996; Fuselier et al., 2000; Lauvraud et al., 2005]. If tailward-of-cusp reconnection lasts for an extended period of time (~hours), the captured plasma can convect toward dawn and dusk and enter the magnetotail, where it forms the cold dense plasma sheet (CDPS) (Fujimoto et al., 1996; Terasawa et al., 1997).

Tailward-of-cusp reconnection jets have been observed in-situ by spacecraft in the exterior cusp (Kessel et al., 1996; Phan et al., 2003; Twitty et al., 2004). Frey et al. (2003) utilized global auroral images to show that the footprint of tailward-of-cusp reconnection was observed for several hours during prolonged northward IMF, i.e., tailward-of-cusp reconnection can be active for extended periods of time.

Low-altitude spacecraft traversing the cusp sample field lines connected to a wide region of the magnetopause and observe the effects of magnetopause reconnection. For southward IMF, as the reconnected field lines move poleward away from a low-latitude dayside magnetopause reconnection site, ions with higher energies move along the field line faster and reach the low-altitude cusp at lower latitudes than lower energy ions. The differences in arrival times create an energy-latitude dispersion in the precipitating cusp ions (Rosenbauer et al., 1975; Shelley et al., 1976; Onsager et al., 1995; Pitout et al. 2009; Connor et al., 2012; Burkholder et al., 2025).

When reconnection occurs tailward of the cusp, the reconnected magnetic field lines convect sunward, resulting in the opposite, or "reversed", ion dispersion, with the highest energy ions at high latitudes and lower energy ions at lower latitudes (Burch et al., 1980; Bosqued et al., 1985; 2001; daSilva et al., 2022). While tailward-of-cusp reconnection is believed to occur during northward IMF (Twitty et al., 2004), it has been suggested that $B_X$-dominated (i.e., radial) IMF could also favor this scenario, as seen in models (e.g., Crooker, 1992; Tang et al., 2013; Lu et al., 2021) and in some observational studies (e.g., Petrinec et al., 2016; Pi et al., 2018). However, the response of the magnetosphere to a radial IMF is not yet understood (e.g., Park et al., 2016; Baraka et al., 2021).

The new NASA Tandem Reconnection and Cusp Electrodynamics Reconnaissance Satellites (TRACERS) mission, a dual satellite mission launched on July 23, 2025, is designed to observe the spatial and temporal effects of magnetopause reconnection, such as ion dispersions and convection, in the low-altitude cusp (Miles et al., 2025). In this paper, we present two consecutive TRACERS-2 northern cusp observations. During the first cusp crossing, the IMF was dominated by a positive $B_Z$ component, and during the second crossing, IMF $B_Z$ was small and $B_X$ was the dominant component. Time History of Events and Macroscale Interactions during Substorms (THEMIS)-D observations from the dayside equatorial magnetopause place the TRACERS-2 cusp measurements in a broader context.

The paper is organized as follows. Section 2 provides an overview of the spacecraft, instrumentation and coordinate systems used in this study. The TRACERS-2 cusp observations are described in Section 3. Section 4 focuses on the THEMIS observations. A discussion and a summary follow in Sections 5 and 6.

## 2. Instrumentation

TRACERS is a two-spacecraft NASA SMEX mission in sun-synchronous, polar orbits at an altitude of ~590 km and a 10:30 Local Time of Descending Node (LTDN) (Miles et al., 2025; Petrinec et al., 2025). From August to November 2025, TRACERS-2 routinely collected data in the northern cusp, while the team was attending to an anomaly associated with the other satellite, TRACERS-1. Since December 2025, both TRACERS spacecraft have been collecting routine tandem measurements in the southern cusp.

In this study, we use single-spacecraft TRACERS-2 data from the northern cusp before TRACERS-1 became operative. We use 312 ms resolution ion measurements from ~8 eV to 20,000 eV from the Analyzer for Cusp Ions (ACI) (Fuselier et al., 2025), 50 ms resolution 40-10,000 eV electron measurements from the Analyzer for Cusp Electrons (ACE) (Halekas et al., 2025), 128 samples/s Hz DC electric field measurements from the Electric Field Instrument (EFI) (Bonnell et al., 2025), and 16 samples/s magnetic field measurements from the Fluxgate Magnetometer (MAG) (Strangeway et al., 2025).

We also use THEMIS-D observations (Angelopoulos, 2008) collected during an inbound magnetopause crossing and observations from ARTEMIS P1 (also known as THEMIS-B; Angelopoulos, 2011) in the solar wind. We use fast survey mode electron and ion observations from the Electrostatic Analyzer (ESA) with an energy range from a few eV up to 30 keV for electrons and 25 keV for ions, with a temporal resolution of 3 seconds for THEMIS-D and 4 seconds for ARTEMIS P1 (McFadden et al., 2008), and magnetic field observations from the Fluxgate Magnetometer (FGM) (Auster et al., 2008).

The TRACERS observations are presented in magnetic local time - magnetic latitude (MLT-MLAT) and solar-magnetic (SM) coordinates where Z is parallel to the Earth's magnetic dipole (pointing north), Y is perpendicular to the Earth-Sun line (positive in the direction opposite to the Earth's orbital motion), and X completes the right-handed system. THEMIS and ARTEMIS data are presented in Geocentric Solar Magnetic (GSM) coordinates.

## 3. TRACERS observations

In this section, we present two consecutive TRACERS-2 crossings of the northern cusp near the peak of the September 30, 2025 geomagnetic storm, when the Disturbance Storm Time (Dst) geomagnetic index reached -98 nT. We show that reversed ion dispersions indicating tailward-of-cusp reconnection were observed in both crossings, despite the different IMF conditions.

### 3.1. IMF conditions

ARTEMIS P1 monitored the solar wind at $(XYZ)_{GSE}$=(-5.2, 60.7, -5.4) $R_E$, and Figure 1a shows 3 hours of IMF near the peak of the storm and time-shifted by -615 seconds by comparing the data to THEMIS-D observations in the magnetosheath.

IMF $B_Y$ was positive and IMF $B_X$ negative throughout the 3-hour interval (Figure 1a). Initially, IMF $B_Z$ was negative. After 07:42 UT, IMF $B_Z$ was positive for more than one hour. Between 09:27 UT and 10:00 UT, the IMF $B_Z$ was mostly positive but small (~0-5 nT), and the IMF was dominated by $B_X$ (~10-13 nT).

The two pairs of vertical black lines in Figure 1a mark the times of two consecutive TRACERS-2 cusp crossings, first at 08:14:55–08:16:20 UT, when the average IMF was $(B_X, B_Y, B_Z)_{GSM}$=(-9.5, 7.9, 13.4) nT, then at 09:50:20–09:51:55 UT, when the average IMF was (-13.2, 8.6, 0.51) nT.

The IMF was relatively stable before and after the times marked in Figure 1a. If we instead calculate the average IMF for an interval that includes 2 minutes before and after the TRACERS-2 cusp crossings, we obtain an average IMF of $(B_X, B_Y, B_Z)_{GSM}$=(-9.8, 8.9, 12.3) nT for the first TRACERS-2 crossing and (-12.7, 9.2, 0.68) nT for the second. Thus, small uncertainties in the exact timing will not affect the IMF associated with each cusp crossing in a significant way.

### 3.2. TRACERS-2 cusp Crossing 1 at 08:14:55 – 08:16:20 UT

Figure 1b shows the TRACERS-2 orbit moving southward through the northern polar cap for the 08:10:18-08:17:09 UT interval. TRACERS-2 encountered precipitating ions indicative of the cusp (green/red colors) between (69.2° MLAT, 13.6 MLT) and (66.1° MLAT, 13.0 MLT). The observed $(\mathbf{E}\times\mathbf{B})/B^2$ convection along the track is indicated by the solid black lines and was dawnward and weakly sunward in the cusp (see also Figure 1i).

Figures 1d-i show the detailed TRACERS-2 observation in the cusp. Between 08:14:55 and 08:16:20 UT, TRACERS-2 observed precipitating ~50 eV-3 keV ions (Figure 1d), which are within the range of magnetosheath energies. At the start of the cusp interval, the precipitating ions exhibited a reversed ion dispersion, with the highest energies at higher MLAT and gradually lower energies at decreasing MLAT (Figure 1d). At the end of the cusp interval, a ~10 second interval of "normal" ion dispersion, i.e., with the highest energies at the lowest latitude, was observed. When both normal and reversed ion dispersions are observed in a single cusp satellite crossing, they are often referred to as "U-shaped". The mechanisms behind the U-shaped signatures are still

a matter of debate (e.g., Sandholt et al., 2002; Reiff et al., 1980; Topliss et al., 2000; Xiong et al., 2024).

The precipitating ions were accompanied by enhanced ~40-600 eV electron fluxes primarily directed downward (Figures 1f,g).

Figures 1h and 1i show the DC electric field observed by TRACERS-2 and the $(\mathbf{E}\mathrm{x}\mathbf{B})/B^2$ convection derived from the electric field and the observed DC magnetic field (not shown), in SM coordinates. Enhanced $\mathbf{E}$ and $\mathbf{E}\mathrm{x}\mathbf{B}/B^2$ values were observed throughout the cusp interval, with $E_X$ and $E_Y$ reaching values of -0.41 V/m and -0.090 V/m, respectively (Figure 1h). The associated convection reached $(\mathbf{E}\mathrm{x}\mathbf{B}/B^2)_X$=0.24 km $s^{-1}$ and $(\mathbf{E}\mathrm{x}\mathbf{B}/B^2)_Y$=-1.5 km $s^{-1}$ (Figure 1i), with some higher frequency fluctuations reaching higher values.

Thus, the convection in the low-altitude cusp was weakly sunward, in agreement with tailward-of-cusp reconnection (e.g., Burch et al., 1980), but dominated by a dawnward component, by a factor of 1:6. The dawnward-dominated convection is consistent with the large positive IMF $B_Y$ (Figure 1a) (Potemra et al., 1984; Crooker, 1988), which favors reconnection between the draped IMF and the geomagnetic field on the dusk side tailward of the cusp in the northern hemisphere, and on the dawn side tailward of the cusp in the southern hemisphere (Lavraud et al., 2006). The IMF clock angle (=arctan($B_Y/B_Z$)) during the TRACERS-2 cusp observations was ~30°, i.e., strongly northward but with a duskward component (0° clock angle is defined as pointing due north ($+Z_{GSM}$) and 90° is duskward ($+Y_{GSM}$)). Figure 2m shows a cartoon depicting the approximate tailward-of-cusp reconnection scenario under such IMF orientation. When reconnection occurs tailward of the cusp on the dusk side in the northern hemisphere, reconnected magnetic fields convect toward the dawnside, and TRACERS-2 observed the low-altitude signature of this convection (Figures 1i). The observed strong domination of dawnward convection over sunward convection in the low-altitude cusp could suggest a suppression of sunward convection by the anti-sunward magnetosheath flow.

### 3.3. TRACERS-2 cusp Crossing 2 at 09:50:20 – 09:51:55 UT

The TRACERS-2 orbit track for the 09:47:53–09:54:52 UT interval, 95 minutes after the first cusp crossing, is shown in Figure 1c. Again, TRACERS-2 encountered precipitating ions indicative of the cusp (green color), this time at higher latitudes, between (73.1° MLAT, 15.2 MLT) and (70.9° MLAT, 14.0 MLT) compared to Crossing 1. As in the previous event, the (**E**x**B**)/$B^2$ convection in the cusp (black straight lines) was dawnward and weakly sunward (Figure 1o).

The detailed TRACERS-2 observations of the cusp are presented in Figure 1j-o. Cusp ions were observed between 09:50:20 UT and 09:51:55 UT (Figure 1j). The precipitating ions had energies of ~50 eV-3 keV, i.e., within the range of magnetosheath energies. Similar to the previous crossing, a reversed cusp ion dispersion was observed at the poleward edge of the cusp ion precipitation (Figure 1j).

Precipitating electrons with ~ 40 eV – 600 eV energies accompanied the cusp ions (Figure 1l). However, in this event, discerning the poleward edge of the cusp electron precipitation is complicated by the presence of inverted-V electrons observed just poleward of the cusp ions. Inverted-V electrons are known to occur in the auroral zone, mainly in the pre-midnight sector, but their occurrence can extend to the dayside postnoon region at high latitudes (Partamies et al. 2008). It is possible that TRACERS-2 skimmed the dusk-side auroral zone just before entering the cusp in this dusk-side cusp crossing (Figure 1c).

The cusp was characterized by enhanced DC electric field (Figure 1n) and strongly dawnward convection (Figure 1o), in agreement with the positive IMF $B_Y$, with a smaller sunward component (Figure 1o), similar to the first cusp event. The **E**x**B**/$B^2$ convection reached 0.46 km $s^{-1}$ and -1.6 km $s^{-1}$ in the $X_{SM}$ and $Y_{SM}$ direction, respectively. Thus, the cusp signatures for this event are similar to those in the first cusp crossing, despite the different IMF conditions (Figure 1a).

## 4. THEMIS-D observations at the dayside magnetopause

In this section, we present THEMIS-D observations from the dayside low-latitude magnetopause around the same time as the TRACERS-2 cusp crossings in Figure 1.

### 4.1. Overview of THEMIS-D observations

Figure 2a-h shows THEMIS-D observations from an inbound magnetopause crossing during the 07:55–08:22 UT interval on September 30, 2025. At 08:11:26 UT, THEMIS-D crossed the magnetopause, marked by a $B_X$ reversal from positive to negative and an increase in ion and electron temperature. The spacecraft position at this time was $XYZ_{GSM}$=(3.6, -11.4, 2.0) $R_E$, corresponding to R=12.1 $R_E$, 7.2 hours MLT, and 9.4° MLAT. Thus, THEMIS-D was located in the prenoon sector while TRACERS-2 crossed the cusp in the postnoon sector (Figure 1). However, THEMIS-D observations still provide relevant information on the dayside magnetospheric response associated with the low-altitude cusp observations.

THEMIS-D was initially in the magnetosheath, characterized by ion energies between ~ 40 eV and 3 keV (Figure 2b) and 10 - 400 eV electrons (Figures 2c,d), high density (>15 $cm^{-3}$) (Figure 2f), low electron temperature (< 40 eV) (Figure 2h), low (< 200 eV) and anisotropic ion temperature with $T_{i\perp} > T_{i\parallel}$ (Figure 2g), and tailward ($V_X < 0$) and dawnward ($V_Y < 0$) ion flow (Figure 2e). After 08:09:51 UT, there was a gradual transition to lower density and higher temperatures, with $T_i$ eventually reaching 3.3 keV and $T_e$ ~340 eV at 08:18:15 UT, indicating a transition to the hot magnetosphere.

Following the magnetopause crossing (at 08:11:26 UT), THEMIS-D remained inside the magnetosphere for the rest of the interval shown. High-speed flows were observed in the region between the magnetosheath proper and the hot magnetosphere (Figure 2e). However, these flows do not seem to be associated with local magnetopause reconnection. For example, an interval of negative $V_Z$ near the magnetopause at 08:14:44–08:17:45 UT reached a value of -70 km $s^{-1}$ which could be consistent with a reconnection site north of the spacecraft. However, the flow speed is significantly lower than the local Alfvén speed in the magnetosheath of 200 km $s^{-1}$, based on the observed density of 32 $cm^{-3}$ and $B_Z$ of 53 nT at 08:09:30 UT. Furthermore, the electron pitch angle distributions indicate closed field lines (e.g., Figure 2l) which is not consistent with a local magnetopause reconnection jet.

**4.2. THEMIS-D observations during the first TRACERS-2 cusp crossing**

While still in the magnetosheath, shortly after 08:00 UT (Figures 2c,d, red dashed vertical line), THEMIS-D observed uni-directional high energy electrons up to 25 keV at 180° pitch angle, not seen at 0° pitch angles. An example of the detailed electron pitch angle versus energy distribution and electron energy fluxes at 0°, 90°, and 180° versus energy is presented in Figure 2k for the 08:01:01–08:01:04 UT measurement. The high energy electron energy flux was enhanced between 100° and 180° pitch angle. The local magnetosheath magnetic field was northward (Figure 2a), thus, the direction of the electrons is consistent with high energy electrons streaming along open field lines away from a reconnection site tailward of the northern high-latitude cusp (Onsager et al., 2001; Lavraud et al., 2005).

Closer to the magnetopause, in the 08:07-08:09 UT interval, high energy electrons were observed at both 0° and 180° pitch angles, indicating bi-directional electrons streaming away from tailward-of-cusp reconnection in both hemispheres. The transition from uni-directional to bi-directional electrons is consistent with tailward-of-cusp reconnection first occurring in the northern hemisphere, then in the southern hemisphere (e.g., Fuselier et al., 2012).

The time interval of the first TRACERS-2 cusp observations (08:14:55-08:16:20 UT; Figures 1d-i), is marked in Figure 2a-h with solid black vertical lines. During this time, THEMIS-D was located in a boundary layer-like region in the vicinity of the magnetopause, with density and temperature intermediate between those in the magnetosheath and the magnetosphere proper. Figure 2l shows the pitch angle versus energy for electrons observed at 08:15:55-08:15:58 UT, during the first TRACERS-2 cusp crossing. At this time, THEMIS-D observed counter-streaming electrons with maximum fluxes at ~100 eV, with perfectly balanced fluxes at 0° and 180° at all energies (Figure 2l, lower panel). These observations are consistent with the capture of heated magnetosheath plasma on closed magnetospheric field lines due to tailward-of-cusp reconnection in both hemispheres (Crooker, 1992; Song and Russell, 1992; Fuselier et al., 2000; Lavraud et al., 2006). The density of the captured magnetosheath plasma varied between 6 $cm^{-3}$ and 14 $cm^{-3}$, the average $T_i$ was ~ 500 eV, and the average $T_e$ ~ 85 eV. These values are in reasonable agreement with previously observed captured magnetosheath plasma near the subsolar magnetopause (Øieroset et al., 2008). Thus, during the first TRACERS-2 cusp crossing, THEMIS-D shows evidence for tailward-of-cusp reconnection in both hemispheres.

To further investigate the connection between the boundary layer observed by THEMIS-D and the TRACERS ion and electron precipitation in the cusp, Figure 2i,j shows electron and ion spectra, respectively, observed by THEMIS in the magnetosheath at 07:57:00 UT (green curves, omni-directional), in the boundary layer at 08:15:26 UT (black curves, 0°-30° pitch angles for electrons), and by TRACERS in the cusp, also at 08:15:26 UT (red curves, averaged over 9 samples for electrons, 5 samples for ions).

The magnetosheath electron differential energy flux peaks at 56 eV while the maximum flux of the magnetopause/boundary layer electrons, presumably captured by the reconnection process tailward of both cusps (Figure 2l), is 90 eV, i.e., significant heating has occurred due to reconnection. The precipitating electrons observed by TRACERS in the low-altitude cusp also peak near 90 eV, i.e., similar to the boundary layer.

The ion spectra also show an increase in peak energy flux from the magnetosheath proper (100 eV) to the magnetopause/boundary layer (400 eV) and peak at a few hundred eV for the precipitating cusp ions. Furthermore, the shape of the ion spectra in the boundary layer and the cusp are remarkably similar, indicating that the plasma populations are the same.

The similarities between the low-altitude cusp ion and electron spectra to magnetopause/boundary layer particle spectra are consistent with both plasmas originating from dual-lobe reconnection.

**4.3. THEMIS-D observations during the second TRACERS-2 cusp crossing**

Figure 3a-h shows an expanded view of the THEMIS-D magnetopause crossing. THEMIS-D remained inside the magnetosphere until the second TRACERS-2 cusp event (09:50:20 – 09:51:55 UT), 95 minutes after the first crossing. THEMIS-D observed boundary-layer like plasma and electron pitch angle distributions similar to those observed in Figure 2l, with perfectly balanced counter-streaming heated magnetosheath electrons (Figure 3i), indicating closed field lines. This suggest that dual-lobe reconnection was occurring at the time of second TRACERS-2 cusp crossing as well. However, since more than 1.5 hours have passed since the magnetopause crossing at 08:11:26 UT, we cannot exclude the possibility that this could be a pre-existing boundary layer created at an earlier time.

## 5. Discussion

### 5.1. Reversed ion dispersion and IMF

The reversed ion dispersion in both TRACERS-2 low-altitude cusp crossings (Figures 1d,j) indicates that reconnection occurred tailward of the high-altitude cusp. The equatorward and dominantly dawnward convection supports this interpretation and is consistent with a reconnection site tailward of the cusp in the dusk sector, in agreement with the positive IMF $B_Y$ (Figure 2m), for both events.

Interestingly, the IMF conditions were markedly different for the two events. For the first TRACERS-2 event, IMF was dominated by a large and positive $B_Z$ component, with $(B_X, B_Y, B_Z)_{GSM}$ =(-9.5, 7.9 13.4) nT, but in the second event, IMF $B_Z$ was small and IMF was dominated by a large and negative $B_X$, with $(B_X, B_Y, B_Z)_{GSM}$ =(-13.2, 8.6, 0.51) nT. Thus, the TRACERS-2 observations indicate that reconnection tailward of the cusp occurred for both northward and radial ($B_x$ dominant) IMF and that their cusp signatures are remarkably similar. The TRACERS-2 cusp observations were obtained in the northern cusp, and the observed negative IMF $B_X$ favors reconnection tailward of the cusp in the northern hemisphere (Crooker et al., 1992; Lockwood and Moen, 1999; Øieroset et al., 1997).

When tailward-of-cusp reconnection occurs during northward IMF, the IMF can drape around the frontside magnetopause and reconnect in both hemispheres (Cowley, 1983, Crooker et al., 1992; Song and Russell., 1992; Raeder et al., 1995). A $B_X$-dominated IMF, on the other hand, can reconnect in the favored hemisphere, but in the opposite hemisphere, the magnetic shear tailward of the cusp would be small, and reconnection would be less likely (e.g., Michotte de Welle et al., 2022). However, tailward of the cusp reconnection in one hemisphere and equatorward of the cusp reconnection in the other hemisphere could be possible, at locations where the magnetic shear is large (e.g., Pi et al., 2018; Trattner et al., 2021).

During the first TRACERS-2 crossing, a fortuitous conjunction with THEMIS-D at the dayside magnetopause provides evidence for dual-lobe reconnection. THEMIS-D observed counter-streaming and perfectly balanced electrons with magnetosheath energies, consistent with capture of magnetosheath plasma on closed magnetospheric field lines by tailward-of-cusp reconnection in both hemispheres (Song and Russell, 1992; Raeder et al., 1995; Sandholt et al., 1999; Fuselier et al., 2000; Lauvraud et al., 2005). Furthermore, outside the magnetopause, THEMIS-D observed unidirectional high-energy electrons at 180° pitch angles, consistent with energized electrons from a reconnection site tailward of the northern cusp, suggesting that reconnection occurred first tailward of the northern cusp, then tailward of the southern cusp. The IMF clock angle during the first crossing was ~ 30°, consistent with Trattner et al. (2021) who found that reconnection occurs tailward of the cusp in both hemispheres for clock angles < 50°. Thus, the observations indicate that dual lobe reconnection occurred during the first TRACERS-2 cusp crossing.

During the second TRACERS-2 cusp crossing, 95 minutes later, THEMIS-D also observed closed field lines and captured magnetosheath plasma, However, these observations were made >1.5 hours after the magnetopause crossing, and it is possible that THEMIS-D observed a pre-existing boundary layer created at an earlier time.

**5.2. Unusually low latitude location of the cusp during geomagnetic storm**

The poleward edges of the two low-altitude cusp crossings were observed at 69° and 73° MLAT, respectively. The average cusp location during northward IMF is ~78°- 80° MLAT (Petrinec et al., 2025 and references therein). The unusually low latitude of the cusp observations presented here could be related to the preceding extended (> 3 hours) interval of southward IMF and high solar wind dynamic pressure during the ongoing geomagnetic storm and associated magnetopause erosion (e.g., Zhou et al., 2000; Le et al., 2015).

Between the first and second TRACERS-2 cusp crossing, the poleward edge of the cusp shifted poleward by 4° MLAT, from 69° MLAT to 73° MLAT. This shift is consistent with dual-lobe reconnection, since this process moves the cusp poleward by adding newly closed field lines to the dayside magnetopause.

## 6. Summary

We have presented two TRACERS-2 cusp crossings at an altitude of 590 km and simultaneous THEMIS-D observations at the dayside magnetopause. We here summarize the main findings.

1. In two consecutive cusp crossings, 95 minutes apart, TRACERS-2 observed reversed ion dispersion and sunward and strongly dawnward convection in the cusp, consistent with reconnection occurring at the high-latitude magnetopause tailward of the cusp on the duskside for IMF $B_Y > 0$.

2. Simultaneous THEMIS-D observations at the dayside equatorial magnetopause during the first TRACERS-2 cusp crossing show evidence for capturing and heating of magnetosheath plasma on closed magnetospheric field lines by tailward-of-cusp reconnection in both hemispheres (Song and Russell, 1992; Raeder et al., 1995; Sandholt et al., 1999; Fuselier et al., 2000; Lauvraud et al., 2005).

3. The similar electron and ion energy spectra in the low-altitude cusp and in the magnetopause/boundary layer is consistent with a common source, i.e., magnetosheath plasma heated by reconnection.

4. The different IMF conditions for the two TRACERS-2 cusp crossings indicate that reconnection can occur tailward of the cusp for both northward and $B_X$-dominated IMF conditions and that these distinct IMF geometries produce remarkably similar signatures in the low-altitude cusp.

5. Between the first and second TRACERS-2 cusp crossing, the poleward edge of the cusp shifted poleward by 4° MLAT, consistent with a net addition of newly closed field lines by dual-lobe reconnection to the dayside magnetosphere during this interval.

6. The unusually low latitude of the cusp could be related to the preceding extended (> 3 hours) interval of southward IMF and the ongoing geomagnetic storm on this day (e.g., Zhou et al., 2000; Le et al., 2015).

This study utilized single-spacecraft observations from TRACERS-2 acquired early in the mission, prior to the operation of the second spacecraft. TRACERS-1 and TRACERS-2 are now collecting two-spacecraft measurements, enabling the separation of temporal and spatial variations. Future conjunction studies with THEMIS and MMS at the magnetopause will further elucidate the temporal and spatial variability of reconnection-driven processes and their coupling to the cusp.

**Acknowledgments**

This research was supported by NASA contracts 80GSFC18C0008 and NAS5-02099, and by NSF grant PHY-2409449 and NASA LWS grant 80NSSC20K1781 at UC Berkeley. The authors declare there are no conflicts of interest for this manuscript.

**Availability Statement**

TRACERS data were obtained from the TRACERS Data Center at https://tracers-portal.physics.uiowa.edu/L2/. THEMIS and ARTEMIS data are from the THEMIS/ARTEMIS Data Center at https://themis.ssl.berkeley.edu/data_retrieval.shtml. The disturbance storm time index (Dst) was obtained from the World Data Center in Kyoto at https://wdc.kugi.kyoto-u.ac.jp/dstdir/.

## Figure captions

**Figure 1.** TRACERS-2 observations from two consecutive crossings through the northern cusp on 2025-09-30, showing reversed ion dispersions and sunward and dawnward convection. (a) Time-shifted IMF data from ARTEMIS P1 in the solar wind, with the times of two consecutive TRACERS cusp observations marked by vertical lines. (b,c) MLT-MLAT TRACERS-2 orbit tracks, with total down-going ion differential energy flux indicated and the cusp visible as enhanced precipitation (green/red). The convection based on $((\mathbf{E}\mathrm{x}\mathbf{B})/B^2)_{X,SM}$ and $((\mathbf{E}\mathrm{x}\mathbf{B})/B^2)_{Y,SM}$ is plotted along the orbit (black straight lines), showing dawnward and sunward convection in the cusp for both events. (d,j) down-going ion spectrogram, (e,k) up-going ion spectrogram, (f,l) 0°-10° pitch angle electron spectrogram, (g,m) 170°-180° pitch angle electron spectrogram, (h,n) DC electric field in SM coordinates, (i,o) $(\mathbf{E}\mathrm{x}\mathbf{B})/B^2$.

**Figure 2.** THEMIS-D observations of an inbound magnetopause (blue vertical dashed line) crossing around the time of the first TRACERS-2 cusp crossing in Figure 1 (vertical black solid lines). (a) Magnetic field in GSM, (b) ion spectrogram, (c,d) 150°-180° and 0°-30° pitch angle electron spectrograms, (e) ion velocity, (f) ion density, (g,h) ion and electron temperature, (i,j) electron and ion differential energy flux in the magnetosheath at 07:57:00 UT (green, omni-directional), boundary layer (black, 0°-30° pitch angle electrons) and low-altitude cusp (red, down-going), both at 08:15:26 UT, (k,l) top: electron pitch angle versus energy, bottom: energy flux versus energy for the 08:01:01-08:01:04 UT and 08:15:55-08:15:58 UT observation times marked with red dashed vertical lines in (a-h). (m) Simplified cartoon illustrating magnetic reconnection tailward of the cusp viewed from the Sun for IMF $B_Y>0$.

**Figure 3.** An extended interval of THEMIS-D observations that includes the times of both TRACERS-2 cusp crossings in Figure 1 (vertical black solid lines). (a-h) same as Figure 2a-h, (i) same as 2k,l for the 09:50:28-09:50:31 UT observation time (red dashed vertical lines in (a-h)).

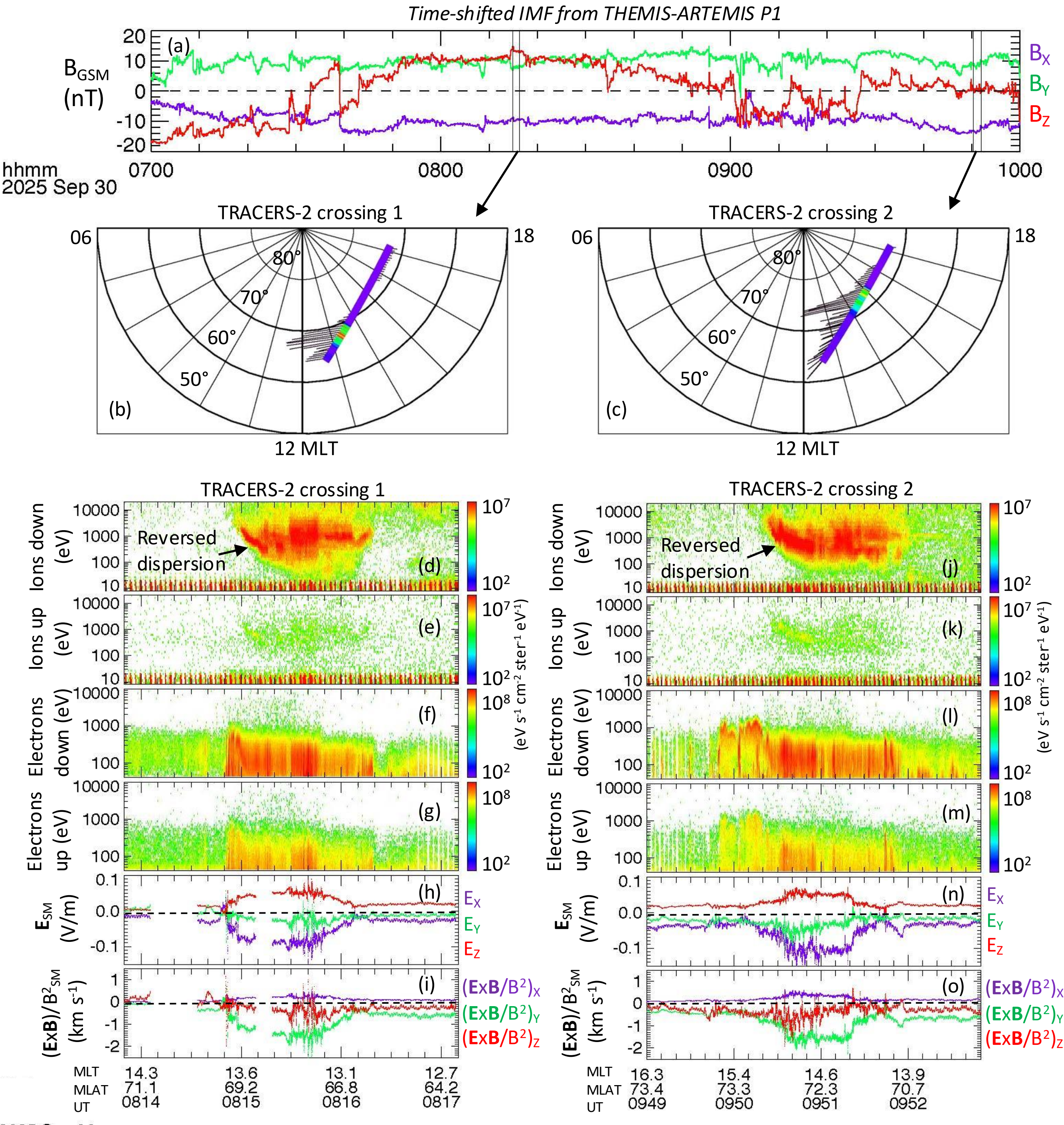

**Figure 1.** TRACERS-2 observations from two consecutive crossings through the northern cusp on 2025-09-30, showing reversed ion dispersions and sunward and dawnward convection. (a) Time-shifted IMF data from ARTEMIS P1 in the solar wind, with the times of two consecutive TRACERS cusp observations marked by vertical lines. (b,c) MLT-MLAT TRACERS-2 orbit tracks, with total down-going ion differential energy flux indicated and the cusp visible as enhanced precipitation (green/red). The convection based on $((\mathbf{E}\mathrm{x}\mathbf{B})/B^2)_{X,SM}$ and $((\mathbf{E}\mathrm{x}\mathbf{B})/B^2)_{Y,SM}$ is plotted along the orbit (black straight lines), showing dawnward and sunward convection in the cusp for both events. (d,j) down-going ion spectrogram, (e,k) up-going ion spectrogram, (f,l) 0°-10° pitch angle electron spectrogram, (g,m) 170°-180° pitch angle electron spectrogram, (h,n) DC electric field in SM coordinates, (i,o) $(\mathbf{E}\mathrm{x}\mathbf{B})/B^2$.

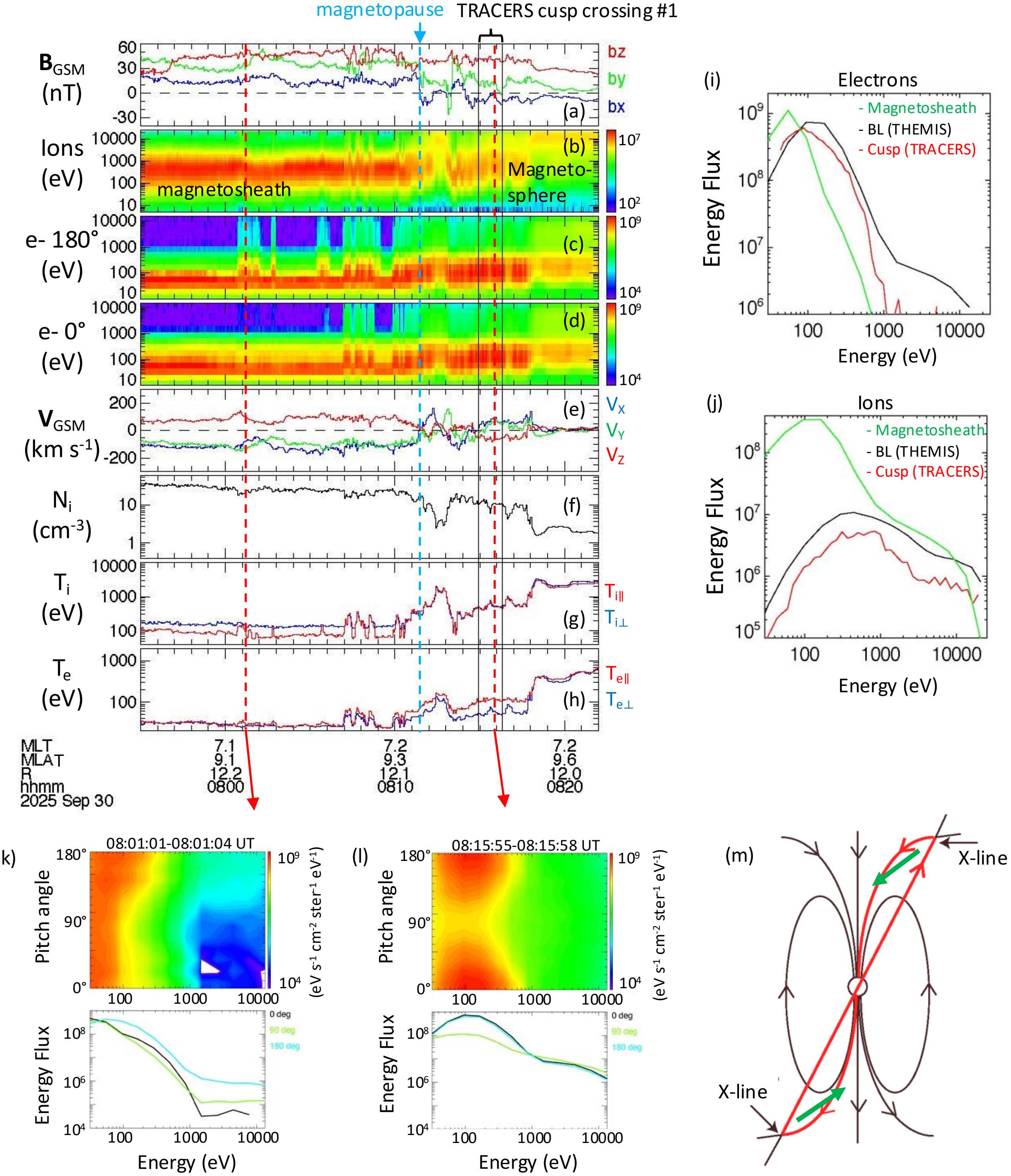

**Figure 2.** THEMIS-D observations of an inbound magnetopause (blue vertical dashed line) crossing around the time of the first TRACERS-2 cusp crossing in Figure 1 (vertical black solid lines). (a) Magnetic field in GSM, (b) ion spectrogram in differential energy flux, (c,d) 150°-180° and 0°-30° pitch angle electron spectrograms, (e) ion velocity, (f) ion density, (g,h) ion and electron temperature, (i,j) electron and ion differential energy flux in the magnetosheath at 07:57:00 UT (green, omni-directional), boundary layer (black, 0°-30° pitch angle electrons) and low-altitude cusp (red, down-going), both at 08:15:26 UT, (k,l) top: electron pitch angle versus energy, bottom: energy flux versus energy for the 08:01:01-08:01:04 UT and 08:15:55-08:15:58 UT observation times marked with red dashed vertical lines in (a-h). (m) Simplified cartoon illustrating magnetic reconnection tailward of the cusp viewed from the Sun for IMF $B_Y > 0$.

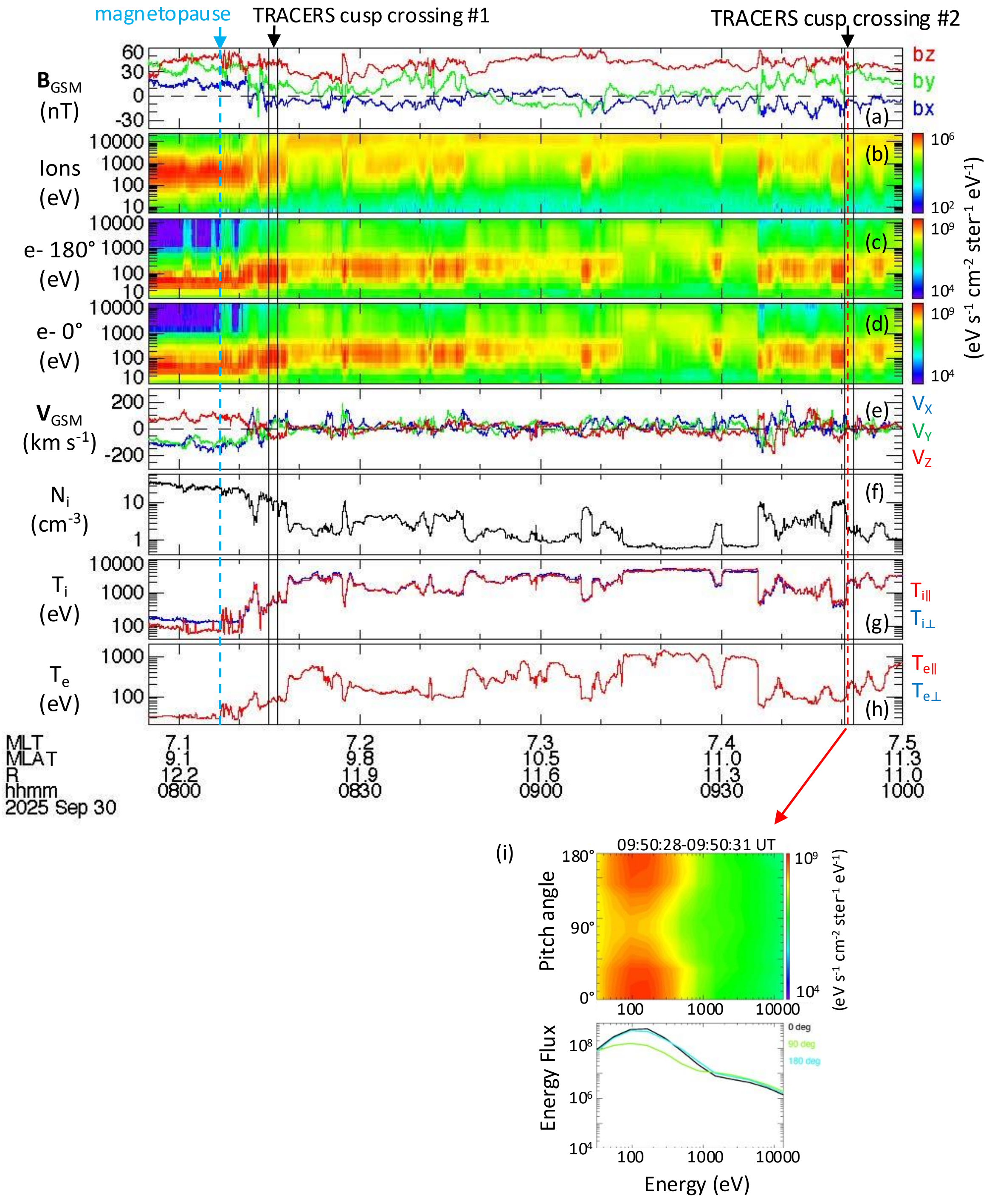

**Figure 3.** An extended interval of THEMIS-D observations that includes the times of both TRACERS-2 cusp crossings in Figure 1 (vertical black solid lines). (a-h) same as Figure 2a-h, (i) same as 2k,l for the 09:50:28-09:50:31 UT observation time (red dashed vertical lines in (a-h)).